\pdfoutput=1
\documentclass[fleqn,usenatbib]{mnras}

\usepackage{newtxtext,newtxmath}

\usepackage[T1]{fontenc}
\usepackage{ae,aecompl}

\usepackage{graphicx}	
\usepackage{amsmath}	
\usepackage{gensymb}
\usepackage{threeparttable}
\usepackage{rotfloat}
\usepackage{ulem}
\usepackage{enumerate}

\title[Cluster AGN Kinematics]{The Impact of Disturbed Galaxy Clusters on the Kinematics of Active Galactic Nuclei}

\author[L. E. Bilton et al.]{
Lawrence E. Bilton,$^{1}$\thanks{E-mail:\href{mailto:l.bilton-2016@hull.ac.uk}{\ l.bilton-2016@hull.ac.uk.}}
Kevin A. Pimbblet,$^{1}$
Yjan A. Gordon,$^{2}$
\\
$^{1}$E.A. Milne Centre for Astrophysics, The University of Hull, Kingston-upon-Hull, HU6 7RX, UK \\
$^{2}$Department of Physics and Astronomy, University of Manitoba, Winnipeg, R3T 2N2 MB, Canada
} 

\date{Accepted XXX. Received YYY; in original form ZZZ}

\pubyear{2020}

\begin{document}
\label{firstpage}
\pagerange{\pageref{firstpage}--\pageref{lastpage}}
\maketitle

\begin{abstract}
We produce a kinematic analysis of AGN-hosting cluster galaxies from a sample 33 galaxy clusters selected using the X-ray Clusters Database (BAX) and populated with galaxies from the Sloan Digital Sky Survey (SDSS) Data Release 8 (DR8).
The 33 galaxy clusters are delimited by their relative intensity of member galaxy substructuring as a proxy to core merging to derive two smaller sub-samples of 8 dynamically active (merging) and 25 dynamically relaxed (non-merging) states.
The AGN were selected for each cluster sub-sample by employing the WHAN diagram to the strict criteria of log$_{10}$([\ion{N}{II}]/H$\alpha$)$\geq-0.32$ and EW$_{\mathrm{H}\alpha}$ $\geq$ 6\AA, providing pools of 70 merging and 225 non-merging AGN sub-populations. 
By co-adding the clusters to their respective dynamical states to improve the signal-to-noise of our AGN sub-populations we find that merging galaxy clusters on average host kinematically active AGN between 0-1.5 $r_{200}$ as $r_{200}\rightarrow0$, where their velocity dispersion profile (VDP) presents a significant deviation from the non-AGN sub-population VDP by $\gtrsim3\sigma$. 
This result is indicative that the AGN-hosting cluster galaxies have recently coalesced onto a common potential.
Further analysis of the composite distributions illustrate non-merging AGN-hosting sub-populations have, on average, already been accreted and predominantly lie within backsplash regions of the projected phase-space.
This suggests merging cluster dynamical states hold relatively younger AGN sub-populations kinematically compared with those found in non-merging cluster dynamical states.
\end{abstract}

\begin{keywords}
galaxies: clusters: general -- galaxies: kinematics and dynamics -- galaxies: active
\end{keywords}



\section{Introduction}
\label{sec:intro}

In a hierarchical universe clustering is inevitable due to the gradual accretion and accumulation of galaxies through successive merger events as a result of gravitational perturbation from the Hubble flow \citep{Regos1989}.
Consequentially, the continued coalescing of galaxies leads to an increase in the likelihood of galaxy-galaxy interactions due to the greater number density of galaxies found at low radii towards the centre of their host galaxy cluster \citep{Moore1996,Moore1999}.
Galaxy clusters are therefore harborers of activity and are found to play host to driving the observed evolutionary differences between cluster and field populations of galaxies \citep{Owers2012}.
These environment-induced gradual dichotomies in galaxy evolution are illustrated through their morphologies, as early-type galaxies become ubiquitous within the densest regions of galaxy groups and clusters, vice versa for late-type galaxies\citep{Oemler1974,Dressler1980,Houghton2015}. 
The trend continues with galaxy colours that typically indicate the average ages of the inhabiting stellar population with redder galaxies, commonly associated with early-type galaxies, lying in regions pertaining to higher number densities \citep{Hogg2003,Hogg2004,Lemaux2019}.
The implication of finding redder galaxies at higher densities is the inference of this correlating negatively with their star formation rates and it is indeed shown that increased density leads to relatively quenched levels of star formation \citep{Gomez2003,Bosch2008,Bamford2009}.
Despite these determined relationships between galactic properties and density they are not the dominant cause for the observed galaxy evolution since field populations are generally mixed, indicative of natural galactic evolution (e.g. see \citealt{Kauffmann2004,Blanton2005,Lemaux2019,Bluck2020}).

The local environment is not purely defined by the greater number densities of cluster galaxies and their interactions with each other however.
There is a diffuse hot gas that pervades the space between the cluster galaxies, the Intracluster Medium (ICM), which has been observed to interact with recently harassed, infalling late-type galaxy populations in particular (e.g. \citealt{Gunn1972}).
As a galaxy gains higher velocities on its passage down into the cluster's deep gravitational potential well, the increasing ICM density will induce ram-pressure stripping of any gas present within the disc and operates on timescales that are inversely proportional to the ICM density (e.g. see \citealt{Gunn1972,Abadi1999,Quilis2000,Roediger2007,Sheen2017}).
If an infalling galaxy experiences continuous ram-pressure stripping the ultimate consequence is the impediment of the star formation processes until quiescence is reached.
The ICM can also interact with an infalling galaxy's own diffuse hot gas halo, which can be easily stripped and, again, result in the premature quenching of star formation processes as their cold gas fuel reservoirs deplete and strangle the galaxy \citep{Larson1980}.

Aside from the atypical intrinsic properties of cluster galaxies that are studied, more recent works investigate the possible connections between the presence of active galactic nuclei (AGN) hosted by cluster galaxies and their local cluster environment.
AGN are themselves a by-product of the accretion of matter into a galaxy's central supermassive black hole, however, not all galaxies possess an active nucleus and this is evident through the observed evolution of quasars as a function of redshift, which peaks at $z\sim2$ similarly to the \cite{Madau1998} plot of star formation history (e.g. see also \citealt{Kauffmann2000,Ellison2011,Kormendy2013}). 
The implications of this signify how AGN must play a role in modulating the growth of stellar mass via some sort of co-evolutionary mechanism, an inference which is strengthened by the strong correlations found between supermassive black hole masses and their host stellar bulge masses (see \citealt{Magorrian1998,Silk1998,Ferrarese2000,Gebhardt2000}).
The transient nature of AGN, albeit on long timescales, is indicative that their `active' nature is dependent on some sort of fuel being accreted onto the central black hole as well as a fuelling mechanism to describe the transport of this fuel.
The mechanisms involved in triggering AGN activity are currently not comprehensively understood, however, it is known the fuel supply is in the form of cold gas that could also contribute to the star forming processes within the host galaxy \citep{Reichard2009}.
As a result reservoirs of cold gas are needed to continually feed the nucleus to make it active, however, the dense regions of galaxy clusters and groups are relatively poor sources of cold gas, although, evidence shows the AGN that do lie within these dense regions are triggered either by cooling gas flows or galaxy-galaxy mergers \citep{Moore1996,Moore1999,Sabater2013}.
One recent revelation for a possible origin of AGN triggering within galaxy clusters is the observed correlation between ram-pressure stripped galaxies--known as `jellyfish galaxies'--and the presence of an AGN residing within these galaxies, implying that the stripped material of an infalling galaxy can cause a migration of fresh cold gas to its supermassive black hole (e.g. \citealt{Poggianti2017,Marshall2018}).  
However, jellyfish galaxies are prevalent in the cores of galaxy cluster\citep{Jaffe2018}, whereas AGN-hosting cluster galaxies that are found to preferentially lie within infall regions \citep{Haines2012,Pimbblet2013}.
This corresponds to the reduction in AGN fraction suggesting that AGN are more likely to become quenched in core regions compared to the infall regions \citep{Pimbblet2012}.
The AGN reduction seemingly continues to operate across group scales with \cite{Gordon2018a} showing a consistent dichotomy in AGN fractions between virialised and infalling regions for group masses log$_{10}$($M_{200}$/M$_{\odot}$)$\geq13$.

Galaxy clusters themselves have less than peaceful histories, with many examples examples of sub-cluster merging processes through interactions in the ICM, the formation of cold fronts and the sub-structuring of the cluster galaxies (e.g. \citealt{Dressler1988,Markevitch2002,Ghizzardi2010,Owers2011,Owers2012,Caglar2017}).
The dynamical states of galaxy clusters can consequently imprint these merger events through their cluster galaxy membership as demonstrated with the aforementioned sub-structuring and grouping of cluster galaxies.
Tests for determining the degree of sub-structuring, such as that of \cite{Dressler1988}, can be used as proxies for delineating between `merging' and `non-merging' cluster environments.
Analysing the cluster galaxy kinematics of these opposing cluster dynamical states via velocity dispersion profiles (VDPs) and rotational profiles can provide an insight into how cluster galaxies, and their sub-populations, respond kinematically to their environment as a function of radius \citep{Hou2009,Hou2012,Bilton2018,Bilton2019,Morell2020}.
In addition, VDPs themselves can independently act as proxies for determining a merging environment if they depict a rising profile as one increases the clustocentric radius within the virial regions, vice versa for non-merging environments (see \citealt{Menci1996,Hou2009,Bilton2018}).
The AGN activity present within galaxy clusters is found to be commonplace within clusters undergoing merging processes, acting as a repercussion to an increase in ram-pressure stripping as a result of the ICM interactions between two sub-clusters \citep{Miller2003,Sobral2015,Ruggiero2019,Ricarte2020}.
Therefore, AGN-hosting cluster galaxies should have their own unique kinematic response to their local environment, providing two unique VDP and rotational profile 'signatures' corresponding to the two aforementioned dynamical states of merging and non-merging galaxy clusters.

Within this work we seek to test the kinematic response of AGN-hosting cluster galaxies between the aforesaid two galaxy cluster dynamical states via VDPs, which are determined utilising a weighted Gaussian smoothing kernel as outlined by \cite{Hou2009}, and via rotational profiles based upon the work by \cite{Manolopoulou2017} and expanded on in \cite{Bilton2019}.
Thereby allowing for the exploration into whether or not the AGN-hosting cluster galaxy kinematics provide results that correspond to prior studies; AGN activity is predominantly found in infalling galaxies while being encompassed by a merging cluster environment.
This is accomplished through obtaining archival galaxy data from the Sloan Digital Sky Survey (SDSS; \citealt{York2000}) in which to build a sample of clusters as defined by X-ray parameters with an X-ray catalogue.
These data and the methodologies in the way they are procured and handled is elaborated within Section \ref{sec:data}.
The computation and output of the AGN kinematics with the VDPs and rotational profiles are detailed in Section \ref{sec:AGNkin}.
Which is followed by discussing the interpretation of the cluster galaxy AGN kinematics in Section \ref{sec:AGNint}.
Concluding with a discussion and summary of the results presented throughout the body of this work in Section \ref{sec:conc}.

Throughout the work presented here we assume a $\Lambda$CDM model of cosmology with $\Omega_{\mathrm{M}} = 0.3$, $\Omega_{\Lambda} = 0.7$, $H_{0} =$ $100h$ km s$^{-1}$ Mpc$^{-1}$, where $h = 0.7$.

\section{The Data}
\label{sec:data}

We briefly outline the methods involved in the procurement and handling of the data used in order to conduct the aims of this work, which follows the same procedures--as well as providing the same cluster sample--used in \cite{Bilton2019}.
This process involves utilising the X-ray Galaxy Clusters Database (BAX; \citealt{Sadat2004}) to collate a list of X-ray clusters that is constrained through parameters defined by the authors.
The respective coordinates for each galaxy cluster that meet the applied parameter limits are then cross-matched with galaxies from SDSS Data Release 8 (DR8; \citealt{Aihara2011}) to build their cluster galaxy memberships.
To provide a definition for our AGN-hosting cluster galaxies, these DR8 galaxies include the $\sim9,400$ deg$^{2}$ of spectroscopy with a magnitude depth of $m_{r} \lesssim 17.7 \ \mathrm{mag}$ in the $r$-band \citep{Strauss2002}.
Specifically, the DR8 spectra were built from the SDSS spectrograph that was comprised of 640 fibres per plate, with each fibre matching to objects on the focal plane of the sky and which are visible to the SDSS.
The spectral resolution ranges from $\lambda/\Delta\lambda=1500-2500$ for the wavelength range of $\lambda=3800$\AA$-9000$\AA.
Additionally, stellar mass estimates from the MPA-JHU value added catalogue are cross-matched with the cluster galaxies, which are used in order to maintain completeness of the sample\citep{Kauffmann2003,Salim2007}.

\subsection{The cluster sample and their cluster galaxies}
\label{ssec:sample}

Utilising the X-ray BAX catalogue we parameterised our sample of clusters to lie within the redshift range $0.0 \leq z \leq 0.15$ to obtain a varied selection of clusters at different epochs of dynamical evolution, while not going too deep so as to impact on the cluster galaxy numbers in order to maintain completeness.
We further constrain our cluster sample by considering only clusters the X-ray luminosity range $1<L_{X}\leq20$ $\times10^{44}$ ergs$^{-1}$ so we select the most massive clusters, resulting in a pool of 431 galaxy clusters.
The DR8 galaxies are matched to their galaxy cluster environments with an initial $\pm0.01\ z$-space and a $\leq10$Mpc $h^{-1}$ projected radius cut from the their respective clustocentric coordinates on the plane of the sky; each cluster galaxy candidate's projected radius is scaled from the BAX-defined galaxy cluster redshifts relative to our pre-defined flat cosmology.
The key global cluster properties of mean recession velocity ($\overline{cz}_{\mathrm{glob}}$) and velocity dispersion ($\sigma_{\mathrm{glob}}$) are calculated for each cluster for cluster galaxies that lie $\leq1.5$ Mpc $h^{-1}$ from their cluster centres.
The velocity dispersions are deduced using the more robust square-root of the biweight mid-variance as defined by \cite{Beers1990}.
The uncertainties for the mean recession velocity and velocity dispersion values are derived following the methodology of \cite{Danese1980}.
We normalise the cluster galaxy redshifts to their respective galaxy cluster mean recession velocities, which is defined as 

\begin{equation}
    \Delta V = c \left (\frac{z_{\text{gal}}-z_{\text{clu}}}{1+z_{\text{clu}}}\right),
    \label{eq:dv}
\end{equation}

\noindent
where we apply a rather restrained upper limit on the velocity around the cluster mean to $\Delta V=\pm1500$ kms$^{-1}$ to mitigate against high likelihood of interlopers.
To define the cluster galaxy membership we deduce phase-space surface caustic profiles using the methodologies of \cite{Diaferio1997,Diaferio1999}, which provide an enclosed trumpet-shaped density profile as a function of the projected radius $R$ for each cluster, thereby formalising the galaxy cluster membership to those galaxies confined within these caustic profiles  \citep{Gifford2013,Gifford2013a}. 
The consequence of these density profiles, where the density evolves as $\rho(r) = 3M(r)/4\pi r^{3}$, is in the computation of the $r_{200}$ and $M_{200}$ that correspond to the values of clustrocentric projected radius and cluster mass where $\rho(r)=200\rho_{c}$, where $\rho_{c}=3H_{0}^{2}/8\pi G$ is the critical density of the flat Universe previously defined. 
Therefore, throughout this work we assume the virial radius of each cluster, which is deemed to be the radial point of virial equilibrium that lies in between galaxies collapsed onto a cluster potential with those that are infalling and beyond, to be approximately $R_{\mathrm{vir}}\sim r_{200}$.

Since we have a sample of clusters across varying redshifts of $z \leq0.15$ we need to be considerate of the sample of available cluster galaxies and maintain completeness in order to mitigate against the Malmquist bias \citep{Malmquist1925}.
We therefore find our cluster galaxies to be complete for those that possess stellar masses of $\mathrm{log}_{10}(M_{\ast}/\mathrm{M_{\odot}})\geq10.2$.
The final steps in the curation of the cluster sample involve simple sanity checks against the interlacing between large-scale structures and whether the galaxy clusters themselves are enriched with enough galaxies for analysis; the \cite{Einasto2001} catalogue was cross-matched to the preliminary cluster sample to help remove known closely-spaced cluster-cluster environments in addition to maintaining a high cluster galaxy richness with the omission of $N_{2.5 r_{200}}<50$ galaxies, where $N_{2.5 r_{200}}$ is the number of galaxies at $< 2.5 r_{200}$.
These procedures lend to a total of 33 galaxy clusters in our sample.

\subsection{Delineating between merging and non-merging galaxy clusters}
\label{ssec:ds}

In order to increase the signal-to-noise of our kinematic analysis between the merging and non-merging dynamical states we will stack cluster galaxies, which are normalised to their respective $\Delta V$ (as per equation \ref{eq:dv}) and $r_{200}$ values, into two sub-samples according to their host galaxy cluster's dynamical state.
However, we first need to establish what we consider to be a `merging' (dynamically active or relaxing) or `non-merging' (dynamically inactive or relaxed) galaxy cluster.
If we are to assume that those galaxy clusters currently undergoing merging processes increase the likelihood of their member cluster galaxies to interact with one another, then one could infer the presence of cluster merging through tracing the intensity of galaxy-galaxy interactions within each cluster.
We therefore implement the \cite{Dressler1988} statistical test for sub-structure ($\Delta$-test) to determine the strength of these galaxy-galaxy interactions as our proxy for determining if a cluster is indeed a merging system.
The $\Delta$-test we employ here compares the differences between the local mean ($\overline{cz}_{\mathrm{local}}$) and local velocity dispersion ($\sigma_{\mathrm{local}}$) with their global counterparts that are calculated for galaxies $\leq1.5$ Mpc $h^{-1}$ from the cluster centre (see equation \ref{eq:ds}).
The local values are computed for each galaxy and its $N_{\mathrm{nn}}=\sqrt{N_{\mathrm{glob}}}$ nearest neighbours, where $N_{\mathrm{glob}}$ is the number of galaxies that lie $\leq1.5$ Mpc $h^{-1}$.

\begin{equation}
    \delta^{2}_{i} = \left(\frac{N_{\mathrm{nn}}+1}{\sigma_{\text{glob}}^{2}}\right) [(\overline{cz}_{\text{local}} - \overline{cz}_{\text{glob}})^{2} + (\sigma_{\text{local}} - \sigma_{\text{glob}})^{2}],
	\label{eq:ds}
\end{equation}

\noindent
where $\delta_{i}$ represents the deviations between the local and global values for a single galaxy and is iterated through for each galaxy $\leq1.5$ Mpc $h^{-1}$ to produce the sum $\Delta=\sum_{i}\delta_{i}$.

The $\Delta$-test is found to be very sensitive in determining the presence of substructuring amongst galaxies and its significance can be found at $\geq99$ per cent when weighted against $N_{\mathrm{MC}}$ Monte Carlo velocity reshuffles \citep{Pinkney1996}.
Therefore, we apply the $\Delta$-test to our cluster sample where substructure is determined to be present at $P\leq0.01$ with our observational $\Delta_{\mathrm{obs}}$ weighted against 1000 Monte Carlo velocity reshuffle simulations $\Delta_{\mathrm{MC}}$.
Where the value of $P$ is computed from the frequency, $f_{\mathrm{MC}}$, in which the condition $\Delta_{\mathrm{obs}} < \Delta_{\mathrm{MC}}$ is met to give $P=f_{\mathrm{MC}}/N_{\mathrm{MC}}$.
This results in two sub-samples of clusters, that are originally defined within \cite{Bilton2019}, that represent our merging and non-merging dynamical states that hold 8 and 25 clusters respectively.
These clusters and their basic properties, including their $\Delta$-test $P$-values, can be found categorised by their dynamical states within Table \ref{tab:bax}.




\begin{table*}
	\centering
	\caption{The mass-complete BAX cluster sample. The J2000 coordinates and X-ray luminosity values are procured from the literature via BAX. The velocity dispersion at $r_{200}$, $\sigma_{r_{200}}$, is determined from the square-root of the biweight midvariance (Beers et al., 1990). The uncertainties for $\sigma_{r_{200}}$ and $\overline{cz}_{\text{glob}}$ are determined using Danese et al. (1980). The values for $N_{r_{200}}$ and $N_{\mathrm{AGN}}$ are the number of galaxies at $\leq r_{200}$ and the total number of AGN at all radii respectively, and are determined for where MPA-JHU \texttt{galSpec} lines have a $\mathrm{SNR} \geq3$, as detailed in section \ref{ssec:whan}. The $P(\Delta)$ values represent the significance of sub-structuring with respect to the $\Delta$-test in equation \ref{eq:ds}. Where $P(\Delta)\ll$0.01 depicts a cluster possessing strong sub-structuring with values smaller three d.p.}
	\label{tab:bax}
	\renewcommand{\arraystretch}{1.2}
	\begin{tabular}{lcccccccr} 
		\hline
		Cluster & RA  & DEC & $L_{x}$ & $\overline{cz}_{\text{glob}}$ & $N_{r_{200}}$ & $\sigma_{r_{200}}$ & $N_{\mathrm{AGN}}$ & $P(\Delta)$ \\
				 & [J2000] & [J2000] & [$\times 10^{44}$ erg s$^{-1}$] & [km s$^{-1}$] & & [km s$^{-1}$] &  & \\
		\hline
		 \multicolumn{9}{c}{Merging}\\
		\hline
		
		Abell 426 & 03 19 47.20 & +41 30 47 & 15.34$^{a}$ & 5396$\pm$62 & 82 & 831$^{+40}_{-46}$ & 5 & 0.010 \\
		
		Abell 1552 & 12 29 50.01 & +00 46 58 & 1.09$^{d}$ & 25782$\pm$111 & 38 & 809$^{+64}_{-84}$ & 8 & 0.003 \\
		
		Abell 1750 & 13 30 49.94 & -01 52 22 & 3.19$^{c}$ & 25482$\pm$95 & 21 & 726$^{+55}_{-71}$ & 9 & $\ll$0.01 \\
		
		Abell 1767 & 13 36 00.33 & +03 56 51 & 2.43$^{c}$ & 20985$\pm$78 & 40 & 770$^{+47}_{-58}$ & 6 & 0.002 \\
		
		Abell 1991 & 14 54 30.22 & +01 14 31 & 1.42$^{d}$ & 17687$\pm$61 & 31 & 535$^{+37}_{-47}$ & 5 & $\ll$0.01 \\
		
		Abell 2033 & 15 11 28.19 & +00 25 27 & 2.56$^{b}$ & 24582$\pm$90 & 17 & 589$^{+51}_{-69}$ & 7 & $\ll$0.01 \\
		
		Abell 2147 & 16 02 17.17 & +01 03 35 & 2.87$^{a}$ & 10492$\pm$48 & 38 & 688$^{+30}_{-35}$ & 15 & $\ll$0.01 \\
		
		Abell 2255 & 17 12 31.05 & +64 05 33 & 5.54$^{a}$ & 24283$\pm$107 & 43 & 817$^{+62}_{-80}$ & 11 & $\ll$0.01 \\
		
		\hline
		 \multicolumn{9}{c}{Non-Merging} \\
		\hline
		
		Abell 85 & 00 41 37.81 & -09 20 33 & 9.41$^{a}$ & 16488$\pm$73 & 28 & 709$^{+44}_{-55}$ & 3 & 0.853 \\
		
		Abell 119 & 00 56 21.37 & -01 15 46 & 3.30$^{a}$ & 13190$\pm$77 & 25 & 760$^{+47}_{-58}$ & 12 & 0.579 \\
		
		Abell 602 & 07 53 19.02 & +01 57 25 & 1.12$^{b}$ & 18587$\pm$94 & 21 & 626$^{+55}_{-75}$ & 8 & 0.163 \\
		
		Abell 1066 & 10 39 23.92 & +00 20 41 & 1.20$^{c}$ & 20985$\pm$91 & 16 & 714$^{+53}_{-69}$ & 5 & 0.020 \\
		
		Abell 1190 & 11 11 46.22 & +02 43 23 & 1.75$^{d}$ & 22484$\pm$87 & 24 & 669$^{+51}_{-66}$ & 13 & 0.194 \\
		
		Abell 1205 & 11 13 22.39 & +00 10 03 & 1.77$^{c}$ & 22784$\pm$106 & 23 & 748$^{+61}_{-82}$ & 7 & 0.026 \\
		
		Abell 1367 & 11 44 29.53 & +01 19 21 & 1.25$^{a}$ & 6595$\pm$49 & 29 & 660$^{+31}_{-37}$ & 3 & 0.026 \\
        
        Abell 1589 & 12 41 35.79 & +01 14 22 & 1.53$^{e}$ & 21585$\pm$88 & 30 & 751$^{+52}_{-66}$ & 7 & 0.124 \\
		
		Abell 1650 & 12 58 46.20 & -01 45 11 & 6.99$^{a}$ & 25182$\pm$100 & 23 & 670$^{+57}_{-77}$ & 10 & 0.636 \\
		
		Abell 1656 & 12 59 48.73 & +27 58 50 & 7.77$^{a}$ & 6895$\pm$40 & 62 & 817$^{+26}_{-29}$ & 6 & 0.087 \\
		
		Abell 1668 & 13 03 51.41 & +01 17 04 & 1.71$^{d}$ & 18886$\pm$89 & 21 & 639$^{+52}_{-69}$ & 9 & 0.336 \\
		
		Abell 1773 & 13 42 08.59 & +00 08 59 & 1.37$^{c}$ & 22784$\pm$96 & 29 & 687$^{+55}_{-73}$ & 7 & 0.336 \\
		
		Abell 1795 & 13 49 00.52 & +26 35 06 & 10.26$^{a}$ & 18587$\pm$92 & 21 & 785$^{+55}_{-69}$ & 4 & 0.265 \\
		
		Abell 1809 & 13 53 06.40 & +00 20 36 & 1.69$^{e}$ & 23683$\pm$80 & 20 & 618$^{+46}_{-60}$ & 5 & 0.420 \\
		
		Abell 2029 & 15 10 58.70 & +05 45 42 & 17.44$^{a}$ & 23084$\pm$102 & 48 & 893$^{+60}_{-76}$ & 15 & 0.415 \\
		
		Abell 2052 & 15 16 45.51 & +00 28 00 & 2.52$^{a}$ & 10492$\pm$65 & 14 & 619$^{+40}_{-50}$ & 4 & 0.663 \\
        
		Abell 2061 & 15 21 15.31 & +30 39 16 & 4.85$^{f}$ & 23383$\pm$69 & 37 & 630$^{+41}_{-51}$ & 11 & 0.183 \\
		
		Abell 2063 & 15 23 01.87 & +00 34 34 & 2.19$^{a}$ & 10492$\pm$78 & 29 & 785$^{+48}_{-59}$ & 8 & 0.016 \\
        
        Abell 2065 & 15 22 42.60 & +27 43 21 & 5.55$^{a}$ & 21884$\pm$98 & 47 & 873$^{+58}_{-73}$ & 15 & 0.211 \\ 
        
        Abell 2069 & 15 23 57.94 & +01 59 34 & 3.45$^{g}$ & 34775$\pm$139 & 23 & 910$^{+77}_{-104}$ & 10 & 0.179 \\
        
        Abell 2107 & 15 39 47.92 & +01 27 05 & 1.41$^{e}$ & 12291$\pm$62 & 17 & 615$^{+38}_{-47}$ & 6 & 0.151 \\
        
        Abell 2124 & 15 44 59.33 & +02 24 15 & 1.66$^{f}$ & 19786$\pm$103 & 17 & 751$^{+60}_{-80}$ & 4 & 0.873 \\
		
		Abell 2199 & 16 28 38.50 & +39 33 60 & 4.09$^{a}$ & 8993$\pm$52 & 30 & 649$^{+33}_{-39}$ & 23 & 0.586 \\
		
        Abell 2670 & 23 54 10.15 & -00 41 37 & 2.28$^{c}$ & 22784$\pm$89 & 42 & 799$^{+53}_{-66}$ & 8 & 0.523 \\
        
		ZWCL1215 & 12 17 41.44 & +03 39 32 & 5.17$^{a}$ & 22484$\pm$86 & 28 & 760$^{+51}_{-64}$ & 6 & 0.873 \\
        
		\hline
	\end{tabular}
	\begin{tablenotes}
		\item $^{a}$ \cite{Reiprich2002} \hspace{0.4cm} $^{e}$ \cite{Jones1999}
		\item $^{b}$ \cite{Ebeling1998} \hspace{1.4cm} $^{f}$ \cite{Marini2004}
		\item $^{c}$ \cite{Popesso2007a} \hspace{1.37cm} $^{g}$ \cite{David1999}
		\item $^{d}$ \cite{Boehringer2000}
	\end{tablenotes}
\end{table*}

\subsection{AGN determination via WHAN diagrams}
\label{ssec:whan}

In order to derive any analysis of AGN-hosting cluster galaxies from our sub-samples we must first define our AGN selection criteria.
Within the confines of optical spectroscopy the selection of AGN has usually been determined by the presence and strength of four narrow emission lines: \ion{H}{$\alpha$}, \ion{H}{$\beta$}, [\ion{N}{II}] $\lambda6584$ and [\ion{O}{III}] $\lambda5007$ as per the diagnostic diagrams of extragalactic spectra by \cite{Baldwin1981}, commonly referred to as `BPT' diagrams. 
However, these BPT diagrams are demanding in requiring all four emission lines to each individually possess a S/N$>3$.
Preserving this condition is indeed important to maintain high quality data with significant results, although, this benefit is negated by the loss of data through sacrificing the completeness of the galaxies sampled.
To be precise, \cite{CidFernandes2010} finds that only $\sim40$ per cent of the emission line galaxies in the region that AGN usually occupy on BPT diagrams will be detected.
\cite{CidFernandes2010} notes a proposition to mitigate against this by reducing the number of narrow emission lines used as a diagnostic for emission line galaxies from four to the two strongest lines, \ion{H}{$\alpha$} and [\ion{N}{II}] $\lambda6584$.

Using these two narrow emission lines AGN can be selected via comparison of the relative strengths of [\ion{N}{II}] $\lambda6584$ and \ion{H}{$\alpha$} with the logarithmic ratio log$_{10}$([\ion{N}{II}]/\ion{H}{$\alpha$}) against the equivalent width of \ion{H}{$\alpha$}, EW$_{\mathrm{H}\alpha}$, in angstroms.
These resultant diagnostics, named as `WHAN' diagrams, define non-passive (i.e. star-forming and AGN dominant) galaxies to lie at EW$_{\mathrm{H}\alpha}>3$\AA \citep{CidFernandes2011}.
In spite of this increase in the completeness of the emission line galaxies there is a complication in the form of contamination of `fake AGN' that would be more appropriately categorised under low-ionisation emission region (LIER), or, star-forming galaxies under the lines of delineation defined by \cite{CidFernandes2011}.
To curb the effects of contamination during the selection of our AGN we opt to use the \cite{Gordon2018a} criteria for the WHAN diagram.
To segregate the star-forming galaxies from the AGN-hosting galaxies a dividing line is placed on the log$_{10}$([\ion{N}{II}]/\ion{H}{$\alpha$}) axis at $-0.32$, thus, denoting galaxies log$_{10}$([\ion{N}{II}]/\ion{H}{$\alpha$})$\geq-0.32$ as AGN and vice versa as non-AGN.
This has been shown to reduce the sample contamination of AGN by star-forming galaxies from $75.88^{+1.06}_{-1.13}$ per cent to $11.07^{+0.99}_{-0.85}$ for \cite{Gordon2018a}.
The other contaminants, LIERs, host weak hydrogen lines and can therefore easily intrude within the `weak AGN' regime defined by \cite{CidFernandes2011} to be 3\AA $\leq$ EW$_{\mathrm{H}\alpha}$ $<$ 6\AA.
Thus, we reduce the contamination of LIERs by adopting the `strong AGN' criteria of \cite{CidFernandes2011} in which we only sample AGN where EW$_{\mathrm{H}\alpha}$ $\geq$ 6\AA.
It is worth noting that during our analysis consideration was made to allay the errors in the stellar mass estimation through the removal of galaxy objects with a significantly broadened Balmer line (see `Broad-line AGN' in \citealt{Gordon2017}), by using the MPA-JHU `\texttt{SIGMA\_BALMER}' velocity dispersions in order to deduce FWHM$_{\mathrm{Balmer}}=2\sqrt{2\ln{2}}\times$[\texttt{SIGMA\_BALMER}], which we were to define by applying a common cut of FWHM$_{\mathrm{Balmer}}$ > 1200 kms$^{-1}$ seen across the literature (e.g. \citealt{Hao2005,Zhang2013,Gordon2017}).
However, we find that the entire MPA-JHU catalogue only yields a maximum FWHM$_{\mathrm{Balmer}} \approx 1177$ kms$^{-1}$ from the `\texttt{SIGMA\_BALMER}' column, which implies prior works that implement this particular cut using MPA-JHU data are doing so fruitlessly.
Furthermore, the accuracy of the stellar mass values is not paramount for the analysis presented here since they are used purely as a proxy of brightness to maintain completeness.

\begin{figure*}
    \centering
    \includegraphics[width=\textwidth]{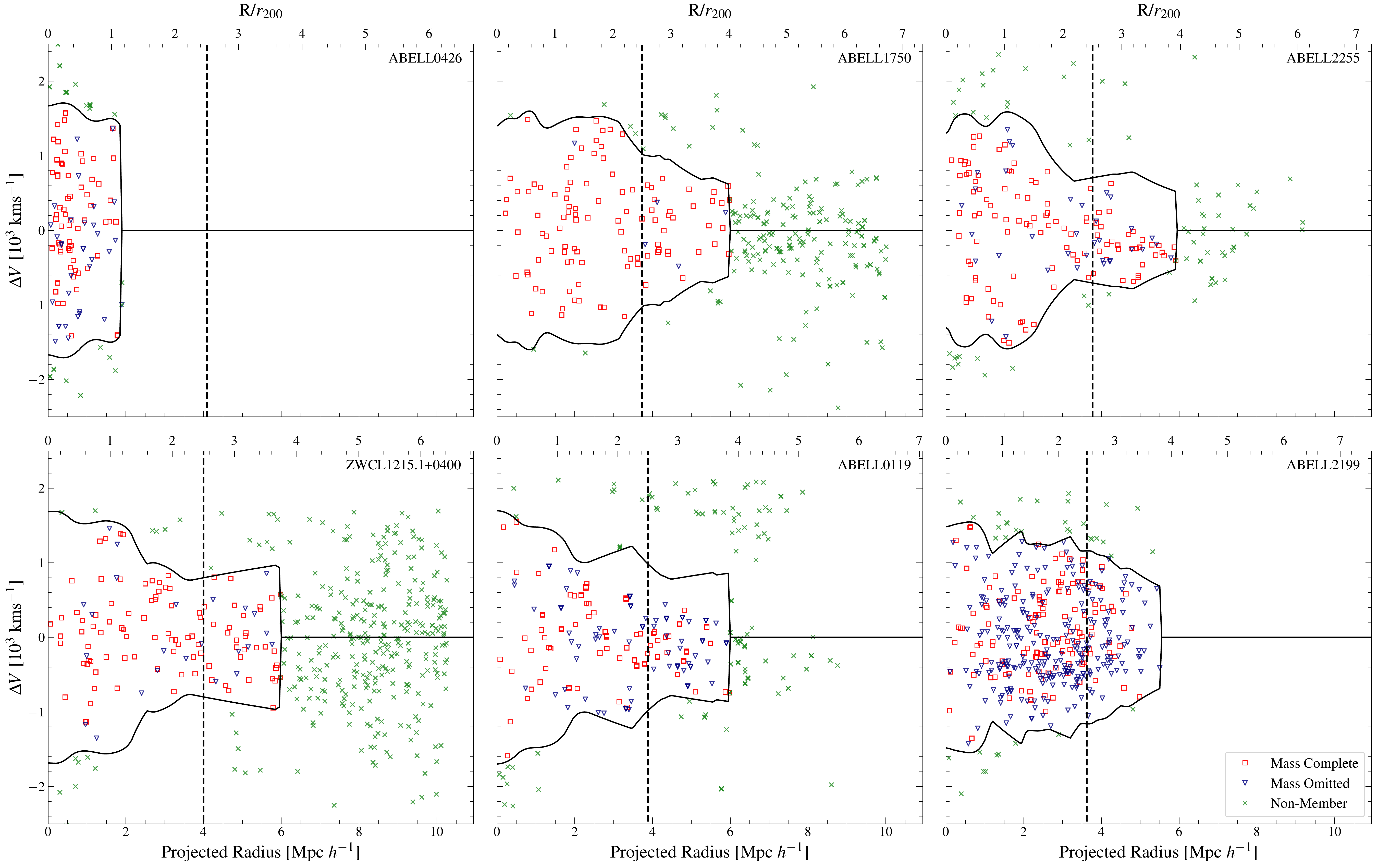}
    \caption{Example phase-space surface caustics (black lines) as a function of the projected radius in units of Mpc $h^{-1}$ to determine the cluster galaxy membership for the merging (top row) and non-merging (bottom row) galaxy clusters in our sample. The hollow red squares indicate the cluster galaxies that are mass complete to log$_{10}(M_{\ast}/\mathrm{M}_{\odot})\geq10.2$, where the hollow blue triangles highlight those cluster galaxies that are omitted (not mass complete) and the green crosses illustrate those cluster galaxies that are not cluster members. The vertical dashed line indicates 2.5$r_{200}$, the upper limit of our kinematic analysis. The cluster galaxy candidates visualised here possess MPA-JHU \texttt{galSpec} lines of $\mathrm{SNR} \geq 3$.}
    \label{fig:caustics}
\end{figure*}

\begin{figure*}
    \centering
    \includegraphics[width=\textwidth]{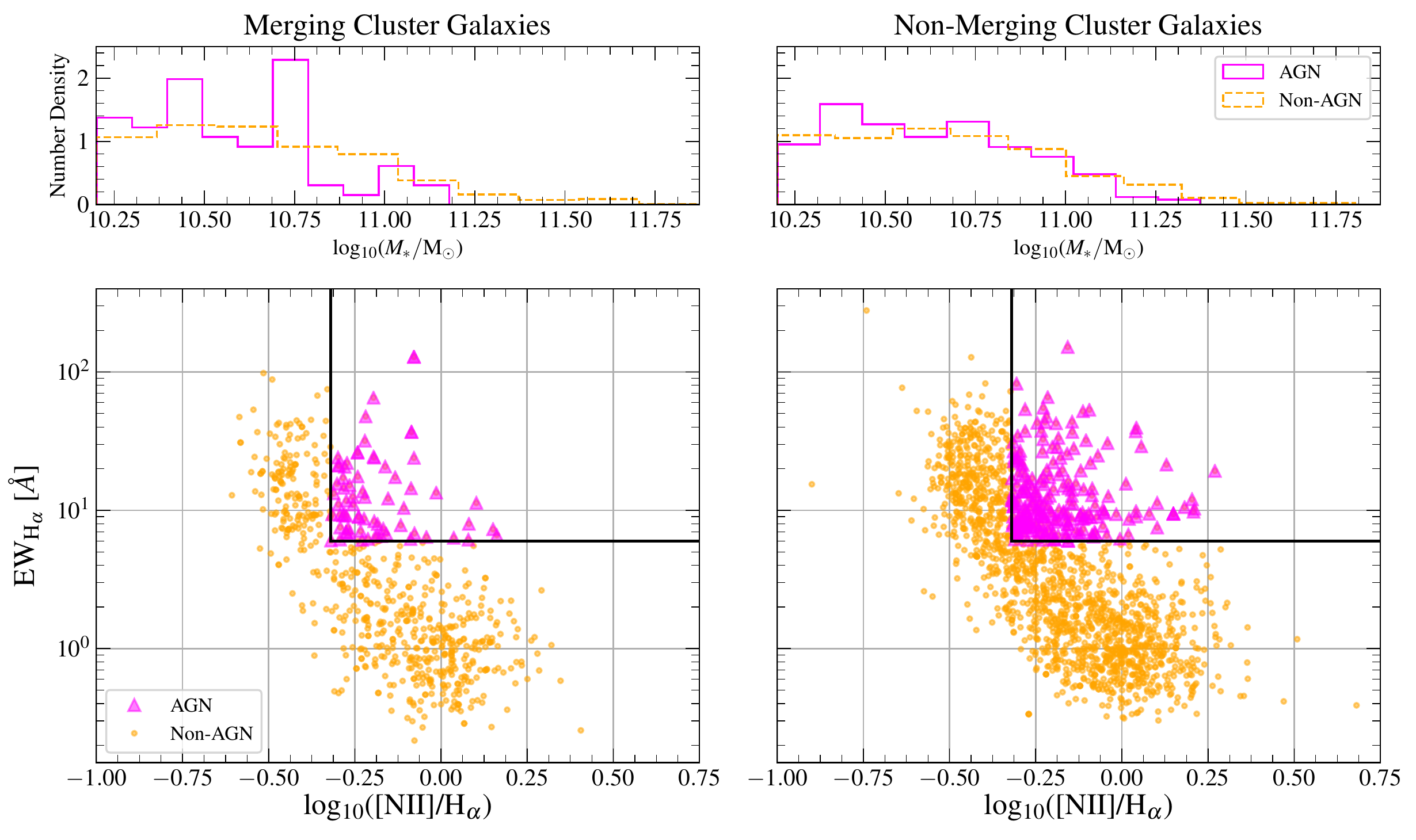}
    \caption{The WHAN diagrams (bottom) for our merging (left) and non-merging (right) sub-samples demonstrate the AGN selection used, with the magenta triangles representing the AGN and the orange dots depicting non-AGN. The thick vertical line represent the ratio of log$_{10}$([\ion{N}{II}]/\ion{H}{$\alpha$})$=-0.32$ and the horizontal lines show the line strength of EW$_{\mathrm{H}\alpha}=$ 6\AA, as per the AGN selection criteria as highlighted in section \ref{ssec:whan}. The distributions of the stellar masses between AGN and non-AGN are also show for their respective sub-samples (top). The frequencies per bin in the stellar mass histograms are normalised to their histogram densities, which is defined as $N = f_{i}/n(c_{i}-c_{i-1})$, where $f_{i}$ is the frequency per bin, $n$ is the total size of the histogram sample and $(c_{i}-c_{i-1})$ is the bin width.}
    \label{fig:whan}
\end{figure*}

As a result of ensuring high levels of completeness and data quality, we sample our AGN sample by maintaining that each narrow line measurement possesses S/N$>3$; we shield against star-forming galaxies by adopting log$_{10}$([\ion{N}{II}]/\ion{H}{$\alpha$})$\geq-0.32$; we maintain stronger ionisation lines to prevent interloper LIER galaxies through enforcing that EW$_{\mathrm{H}\alpha}$ $\geq$ 6\AA.
Applying this to each of the galaxy cluster sub-samples as a whole provides 70 AGN and 686 non-AGN in the merging sub-sample against 225 AGN and 1713 non-AGN in the non-merging sub-sample, providing an AGN fraction of 10.20 and 13.14 per cent of the total cluster galaxies respectively.
An example of the aforementioned surface caustics produced in section \ref{ssec:ds}, which define our cluster galaxy membership from each sub-sample, can be found in Figure \ref{fig:caustics} with cluster galaxies possessing MPA-JHU \texttt{galSpec} lines of $\mathrm{SNR} \geq 3$.
The WHAN diagrams for each stack are shown in Figure \ref{fig:whan} alongside the distributions of the stellar masses for AGN and non-AGN cluster galaxies.

Additionally, we note that the mass distributions between the AGN and non-AGN in the merging dynamical state show a slight deviance from each other.
Therefore, we test whether these distributions are drawn from the same pool of cluster galaxy masses using the two-sampled Kolmogorov-Smirnov test, which yields the p-value $P(\mathrm{KS})=0.027$ and the KS statistic $D_{\mathrm{stat}}=0.187$.
Interestingly, for a significance of $\geq95$ per cent ($P(\mathrm{KS})\leq 0.05$) the two-sampled KS-test indicates a rejection of the null hypothesis with the $D_{\mathrm{crit}}=0.170$, which can be seen in the displacement of the medians between the two distributions with AGN and non-AGN showing $10.52$ log$_{10}$($M_{\ast}/\mathrm{M}_{\odot}$) and $10.62$ log$_{10}$($M_{\ast}/\mathrm{M}_{\odot}$) respectively.

\section{Cluster Galaxy AGN Kinematics}
\label{sec:AGNkin}

\subsection{AGN Velocity Dispersion Profiles}
\label{ssec:AGNVDP}

The kinematics of the AGN are derived for each sub-sample via the computation of VDPs, which are elucidated from the data by normalising their host clusters onto a common phase-space and are thereby co-added according to their pre-defined merging or non-merging dynamical states.
The VDPs we produce in this work are functions of the projected radius, $\sigma_{P}(R)$, originally devised by \cite{Bergond2006} for analysing the kinematics of stellar systems but have since been extended to the large-scale structures of galaxy groups and clusters by a variety of authors (e.g. \citealt{Hou2009,Hou2012,Pimbblet2014,Bilton2018,Morell2020}).
These VDPs are calculated through cluster galaxy radial velocities at fixed incremental bins of radius, with each bin weighted against a Gaussian window function that is driven exponentially by the square of the difference in radius for each $ith$ galaxy.
This window function, corrected by \cite{Bilton2018}, is thus written as

\begin{equation}
	\omega_{i}=\frac{1}{\sigma_{R}}\exp{-\Bigg[\frac{(R-R_{i})^{2}}{2\sigma_{R}^{2}}\Bigg]},
	\label{eq:weight}
\end{equation}

\noindent
where $\sigma_{R}$ is the width of the moving window that weights the window function and $(R-R_{i})^{2}$ is the square of the difference in projected radius.
We set the width of the window to $\sigma_{R}=0.2R_{max}$ in units of $r_{200}$.
Setting the window width to this size allows for us to elucidate the variation in kinematics to a relatively small scale without becoming too fine to the point of inducing a spurious response in the final profile.
Following the calculation of the window function the projected VDP can be deduced, which is written as

\begin{equation}
	\sigma_{P}(R)=\sqrt{\frac{\sum_{i}\omega_{i}(R)(x_{i}-\bar{x})^{2}}{\sum_{i}\omega_{i}(R)}},
	\label{eq:vdp}
\end{equation}

\noindent
where $(x_{i}-\bar{x})^{2}$ is the square of the difference in the radial velocities between the $ith$ galaxy and the mean recession velocity of the cluster.
The result of parsing equation \ref{eq:weight} through equation \ref{eq:vdp} for each bin of radius is a smoothed radial velocity profile that responds to every galaxy and their proximity to the bin.
To maintain the validity of this VDP methodology for analysing the kinematics it is wise to ensure the total number of cluster galaxies used to output a profile meets the lower limit of $20$ members. 
If too few cluster galaxies contribute to the profile this can lead to an unrealistic response due to the weightings that depend on the projected separation between galaxies and fixed bins with the consequence of large uncertainties.

We incorporate the aforementioned systematic processes for each of our cluster sub-samples so as to be able to partly satisfy the aims of this body of work to compare the kinematic response of AGN-hosting cluster galaxies between different galaxy cluster dynamical states.
The procedure we follow for the VDP production is simple, and thus, it outlined here: cluster galaxies are collated from every cluster into their respective merging or non-merging sub-samples as per the definition described in subsection \ref{ssec:ds}.
These cluster galaxies line-of-sight velocities are normalised to their host cluster's mean recession velocities to provide $\Delta V$, which is weighted to the $\sigma_{r_{200}}$, with their projected radii to $r_{200}$ and are co-added onto a common $\Delta V/\sigma_{r_{200}}-r_{200}$ grid to output merging and non-merging phase-space stacks.
After the allocation of the cluster galaxies to their appropriate dynamical states, the AGN selection criteria of subsection \ref{ssec:whan} is applied to ascertain the AGN present for both sub-samples.
Finally, the AGN and non-AGN cluster galaxies for each sub-sample are computed through into equations \ref{eq:weight} and \ref{eq:vdp} to result in a total of four profiles, two for each dynamical state.

\begin{figure*}
    \centering
    \includegraphics[width=0.84\paperwidth]{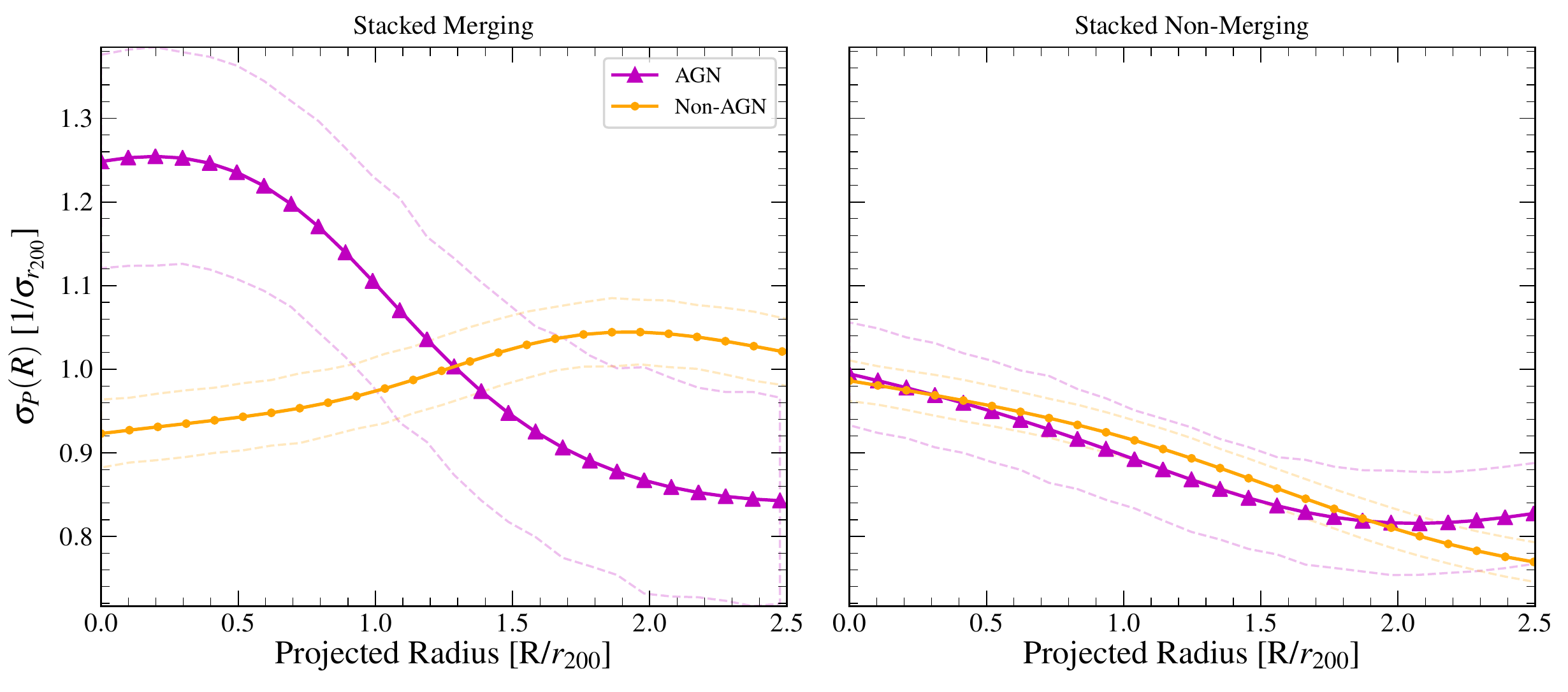}
    \caption{The VDPs split by our AGN selection, for the merging (left) and non-merging (right) dynamical states, produced via co-adding clusters appropriately onto a common phase-space grid. Each stack shows the projected velocity dispersions ($\sigma_{P}{R}$), in normalised units of $1/\sigma_{r_{200}}$, for AGN and non-AGN sub-popluations as a function of radius ($R/r_{200}$); the magenta triangle markers represent the AGN and the orange dot markers non-AGN. The corresponding dashed lines represent the symmetric uncertainty for each profile derived from $1\sigma$ of 1000 Monte Carlo resamples.}
    \label{fig:AGNVDP}
\end{figure*}

We show the product of our VDP implementation for each dynamical state between AGN and non-AGN cluster galaxies in Figure \ref{fig:AGNVDP}.
Firstly, focusing on the non-merging VDPs in the right panel of Figure \ref{fig:AGNVDP}, we witness the AGN and non-AGN profiles declining in parity with one another as the projected radius increases until $R\sim2 \ r_{200}$ where the AGN profile starts to break away and increase.
The near-perfect parity between both of these VDPs suggests that the AGN population has homogenised with the non-AGN population and are not interacting beyond the expected settling of the normalised velocity dispersion to $\sim1$, representing a relaxed stack of galaxy clusters.
This is not an unexpected result considering this sub-sample marries to the non-merging VDPs of \cite{Bilton2018}, where the cluster galaxy sub-populations of stellar mass, galaxy colour and galaxy morphology consistently demonstrate this decline as a result of the relaxed dynamical state (see also \citealt{Girardi1996}).
This is in contrast to the merging VDPs in the left panel of Figure \ref{fig:AGNVDP}, where the AGN-hosting cluster galaxy VDP rises to values of $\sigma_{P}(R) \sim 1.25$ as $R\rightarrow0$, diverging from the non-AGN cluster galaxy VDP with a significance of $\gtrsim3\sigma$ at $R=0$.
As the projected radius extends outward from the clustocentric regions the AGN sub-population steeply declines in their kinematic activity to equivalent levels seen for a non-merging dynamical state.
The increase in the projected velocity dispersion of an AGN sub-population towards the centre of the merging stack implies that these AGN are on their first infall, or, that they are residing within backsplash galaxies (see the VDPs in Figure 13 of \citealt{Haines2015}).
Here backsplash galaxies are recently accreted cluster galaxies that have already passed through their pericentres and are proceeding to journey to their apocentres \citep{Pimbblet2011,More2015,More2016}.
Although, we should highlight that the number of AGN-hosting cluster galaxies lying at $\leq r_{200}$ in the merging cluster stack is only 15 compared to the 55 found $> r_{200}$, which indicates there is a possibility the rise in the VDP is spurious due to inadequate sampling of the AGN.
The non-AGN sub-population, however, illustrates an opposing response where the profile increases steadily with $R$ reaching an apex at $R\sim1.8 \ r_{200}$.
This is, again, an unsurprising result considering prior works have shown merging populations of cluster galaxies possess an rise in their kinematic activity as $R$ increases with the red and blue sub-population VDPs inferring strong sub-clustering along with the presence of `pre-processing' (see \citealt{Menci1996,Hou2009,Bilton2018}).

\subsection{Cluster Galaxy AGN Rotational Profiles}
\label{sec:AGNrot}

Another testable indirect method of determining potential cluster environmental effects that could trigger AGN is analysing the `rotational profiles' of our selected AGN sample between the two dynamical states, which are naturally contrasted against those that are `non-AGN'.
Galaxy clusters themselves are known to possess some sort of global angular momentum that operates dynamically with respect to the bottom of a cluster's potential well (e.g. see \citealt{Materne1983,Oegerle1992,Hwang2007,Manolopoulou2017,Baldi2018}).
Indeed, any angular momentum possessed within a galaxy cluster should influence the average motion of the galaxy cluster membership via these very dynamics, which would be imprinted onto the radial velocities of the individual cluster galaxies in $z$-space.
Thus, following the combined methodologies detailed within \cite{Manolopoulou2017} and \cite{Bilton2019}, we determine the relative rotational profiles of our aforementioned cluster galaxy sub-populations from the 2D plane of sky through the employment of a geometric `perspective rotation' technique \citep{Feast1961}.

Perspective rotation relies upon the projection of 3D motions of cluster galaxies onto a 2D RA-DEC space relative to a known cluster centre.
Thus, with the known BAX defined galaxy cluster coordinates and the known RA and DEC values of each member galaxy one can determine their projected angles with respect to a defined normal.
Furthermore, by artificially rotating the cluster galaxies about their respective BAX centres it is possible to determine the planar angle of rotation through finding the maximum difference between the averaged radial velocities for either side of the defined normal.
We outline our procedure for determining the cluster galaxy sub-population rotational profiles firstly be making the assumption that the rotational axis of each cluster in our sample lies solely in the plane of the sky so they are perpendicular to our line-of-sight, which leaves the angle of the rotational axis perpendicular to the plane $\phi=0\degree$, consequently defining the line-of-sight velocity to be $v_{\mathrm{los}}=\Delta V$ (see \citealt{Manolopoulou2017}).
For each galaxy cluster we generate a fixed normal line along their central declination as defined by the X-ray literature with the BAX catalogue, which allows for the calculation of the cluster galaxy's projected angles with respect to this normal, denoted as $\mu$. 
This fixed normal simultaneously acts as a divide upon which we calculate the averaged $v_{\mathrm{los}}$ for the two semicircles $\langle v_{1} \rangle$ and $\langle v_{2} \rangle$.
These are defined as


\begin{equation}
 	\langle v_{1,2} \rangle = \frac{1}{N} \sum_{i=1}^{N} \Delta V_{i} \cos(90\degree-\mu_{i}),
    \label{eq:vels}
\end{equation}

\noindent
where $\Delta V_{i}$ is the line-of-sight velocity from equation \ref{eq:dv} for the galaxy $z_{\text{gal,}i}$ and $\mu_{i}$ is the angle from the normal operating between $0\degree$ and $180\degree$ for each semicircle.
Using Equation \ref{eq:vels} allows to ascertain the difference in averaged velocities with $v_{diff} = \langle v_{1} \rangle-\langle v_{2} \rangle$ and is, therefore, iterated through rotating the cluster galaxies about their galaxy cluster centre by $\theta=10\degree$ until $\theta=360\degree$.
In addition, we procure the uncertainties of each semicircle by propagating through the standard error for each semicircle at every increment of $\theta$ as

\begin{equation}
    \sigma_{\theta} = \sqrt{\frac{\sigma^{2}_{v,1}}{n_{1}} + \frac{\sigma^{2}_{v,2}}{n_{2}}},
    \label{eq:unc}
\end{equation}

\noindent
where $\sigma_{v}$ is the velocity dispersion and $n$ is the galaxy number for each semicircle 1 and 2 at each increment of $\theta$.

To match our global galaxy cluster property definitions we apply Equation \ref{eq:vels} and \ref{eq:unc} for all clusters across both merging and non-merging sub-samples for their cluster galaxies at a projected radius of $\leq1.5$ Mpc $h^{-1}$.
We thus take the maximum values of $v_{\text{diff}}(\theta)$ for our global definition of the rotational velocities ($v_{\text{glob}}$) for each galaxy cluster, ergo this proceeds to provide the planar rotational axis $\theta_{\text{glob}}$.
The global rotational values and statistics for the sample of galaxy clusters presented within the body of this work are defined and catalogued in full in \cite{Bilton2019}. 

Continuing on from the previously outlined methodology we build two stacks of galaxy clusters from our two sub-samples, where the respective cluster galaxies are co-added onto normalised RA-DEC grids with their X-ray centres set to zero.
This is alongside the cluster galaxy radial velocities, which are derived to their respective mean recession velocities as per Equation \ref{eq:dv} and are normalised by the velocity dispersion $\sigma_{r_{200}}$, similar to the composites produced for subsection \ref{ssec:AGNVDP}.
Additionally, each set of cluster galaxies from each galaxy cluster are rotated about their origin by $\theta_{\text{glob}}$ to align their planar rotational axes along the same normal so as to not overlap opposing dynamics and ensure we enhance the signal of our rotational profiles.
This provides a rotational axis of $\theta\sim0\degree$, thus, implying the maximum value is consistently found at $v_{\text{diff}}(\theta=0)$ as we increase incrementally in $R$ where we define  $v_{\mathrm{rot}}=v_{\text{diff}}(0)$. 
Therefore, with each composite of cluster galaxy sub-populations for each dynamical state we exploit Equation \ref{eq:vels} to determine the $v_{\mathrm{diff}}$ as a function of radius in increments of $0.1 r_{200}$ over $0<R\leq2.5 \ r_{200}$ to maintain consistency.

\begin{figure*}
    \centering
    \includegraphics[width=0.84\paperwidth]{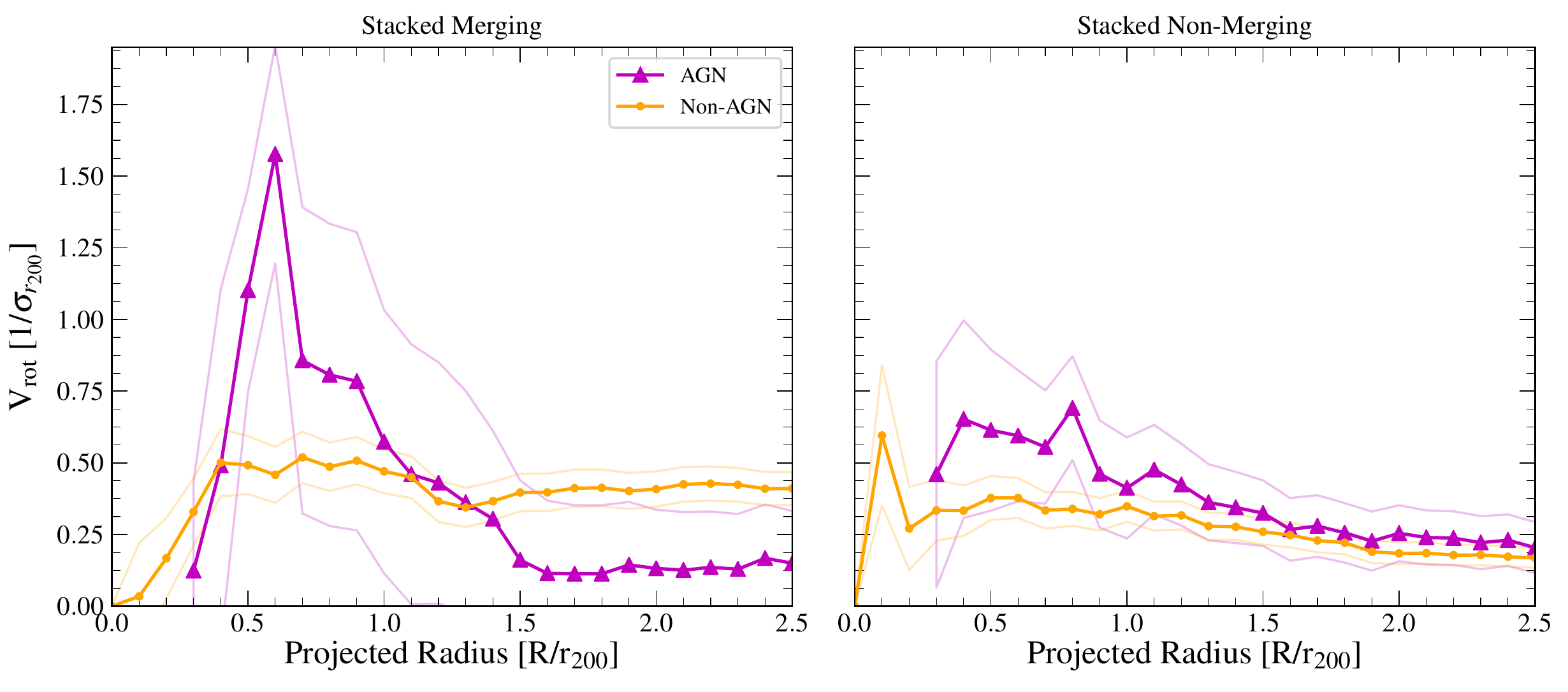}
    \caption{The AGN (magenta triangles) and non-AGN (orange dots) $V_{\mathrm{rot}}$ profiles for the cluster galaxies in the merging (left) and non-merging (right) dynamical states. The respective regions around each of the profiles, as shown with the solid lines, represent the uncertainty obtained via propagated standard errors of the mean as per Equation \ref{eq:unc}.}
    \label{fig:AGNrot}
\end{figure*}

In Figure \ref{fig:AGNrot} we present the rotational profiles of our selected AGN sample contrasted with the non-AGN for the merging and non-merging dynamical states that were defined in subsection \ref{ssec:ds}.
Concentrating on the non-merging sub-populations both rotational profiles show no significant deviation from one another and appear to be homogenised in a similar fashion to the VDPs in Figure \ref{fig:AGNVDP}, with the AGN sub-population lacking detail close to the core regions due to the dwindling numbers that occupy them.
If we consider the rotational profiles from \cite{Bilton2019} we can see the general trend of a relatively quenched and decline profile with radius is consistent despite our strict demand for strong and significant line emissions.
Although, there is no significant discrepancy between the AGN and non-AGN profiles, which coinciding with the stellar masses presented in Figure \ref{fig:whan} suggests the AGN within this sample are drawn from the same distribution as the non-AGN, most likely coalescing onto their cluster potentials simultaneously at the same epochs.  
The co-added merging cluster galaxies almost depict a similar outcome of homogenisation from the analysis, however, the AGN sub-population does briefly spike to a V$_{\mathrm{rot}}\sim1.5$ at $R\sim0.6 r_{200}$ to a significance of $\sim 2\sigma$ from the non-AGN sub-population.
Furthermore, this is followed with a steep declining gradient that flattens at $v_{\mathrm{rot}}\sim0.1$ at $R\gtrsim1.5 r_{200}$.
Overall, the connotations of the observed spike and decline, while noisy, can corroborate that these AGN either contribute to an infalling or backsplash population of cluster galaxies with the merging AGN sub-population VDP in Figure \ref{fig:AGNVDP}.
Although, despite the increased variation in the AGN profile, the large uncertainties and insufficient numbers of AGN that contribute to the merging stack impede one's ability to be conclusive about the kinematic independence of the sub-population relative to the non-AGN profile. 

\section{Interpretations of the AGN Kinematics}
\label{sec:AGNint}

We have thus far presented how AGN-hosting cluster galaxies respond kinematically as a function of projected radius between unrelaxed and relaxed galaxy cluster dynamical states, however, we are yet to explore what the key results presented in Figures \ref{fig:AGNVDP} and \ref{fig:AGNrot} imply about the possible origins of AGN in galaxy clusters based upon prior knowledge and works.
To elaborate, \cite{Poggianti2017} has shown with MUSE spectra that so-called `Jellyfish' galaxies--a cluster galaxy with extended tails of gas and stars as a result of ram pressure stripping with the ICM (e.g.\citealt{Yagi2010,Kenney2014,Rawle2014})--seemingly are more likely to posses and AGN with 5/7 of jellyfish galaxies containing an active nucleus, which is further confirmed with evidence of outflows and ionisation models matching AGN profiles within \cite{Radovich2019}.
Additionally, increased star formation and AGN activity has been found in cluster-cluster mergers and by extension this includes the jellyfish morphologies, which have been consistently found to harbour within merging cluster environments as well, with the more extreme cases being the result of interactions with high velocity cluster merger shock fronts in the ICM \citep{Miller2003,Owers2012,McPartland2016,Ebeling2019}.
However, the Abell 901/2 system of simultaneously interacting two sub-clusters and two sub-groups is one of the more plentiful reservoirs of jellyfish galaxies of 70 and only 5 of these galaxies host an AGN, indicating the mechanisms involved in triggering AGN must depend on more parameters than just the coincidence of jellyfish morphologies \citep{RomanOliveira2019}.
Despite this caveat, the link between ram pressure stripping and an increase in the AGN activities has continued to show promise with simulations by \cite{Ricarte2020}, determining galaxies with a mass log$_{10}(M_{\ast}/\mathrm{M}_{\odot})\gtrsim 9.5$ have spikes in black hole accretion as the star formation is quenched around the strongest regions of ram pressure stripping as the galaxy journeys through its pericentre.
Furthermore, the simulations by \cite{Ricarte2020} seem to illustrate how the quenching of star formation is aided by AGN feedback as a consequence to the spikes on AGN activity and thus producing outflows until the AGN itself runs out of fuel; observational evidence backs this claim of AGN feedback \citep{George2019}.
From this brief overview, our Figures \ref{fig:AGNVDP} and \ref{fig:AGNrot} demonstrate an immediate interpretation that our merging dynamical state represents the AGN sub-population to be hosted by recently accreted cluster galaxies, corroborating the simulations of \cite{Ricarte2020}.
Placing the current established lines of enquiry on the mechanisms that lead to AGN triggering into consideration we attempt to isolate the nature of their host cluster galaxies; Do AGN-hosting cluster galaxies represent a sub-population of galaxies on their first infall, or, are these galaxies representative of a backsplash population to account for the AGN spikes during the passage though their respective pericentres?
We therefore briefly attempt to interpret the VDPs and rotational profiles with complementary analysis, which is detailed in the following sub-section.

\subsection{Backsplash Cluster Galaxies}
\label{ssec:bkspl}

\begin{figure*}
    \centering
    \includegraphics[width=0.8\paperwidth]{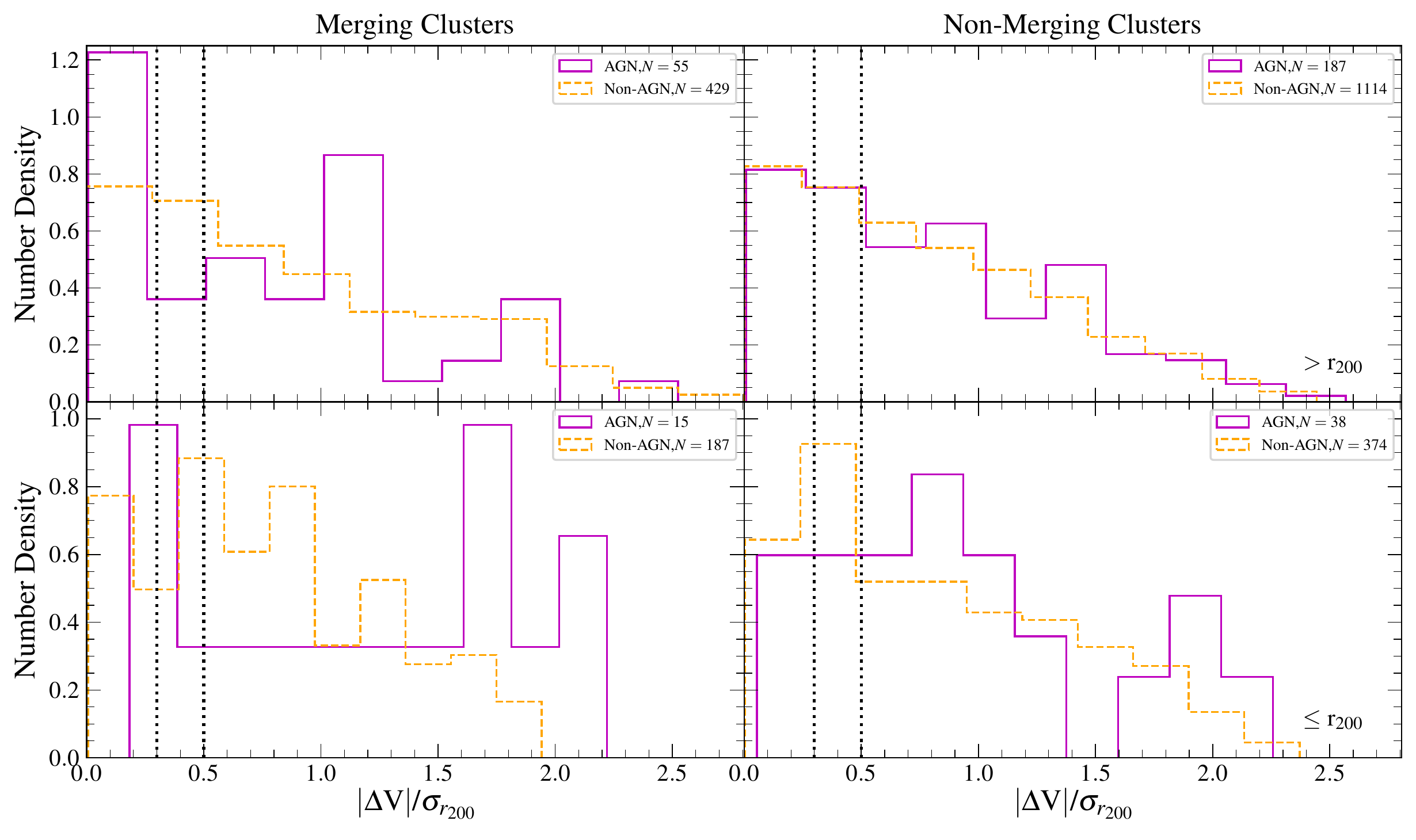}
    \caption{Histograms of $|\Delta\mathrm{V}|/\sigma_{r_{200}}$ for AGN and non-AGN sub-populations segmented into bins of cluster galaxies that lie > $r_{200}$ (top row) and $\leq r_{200}$ (bottom row) between merging (left column) and non-merging (right column) dynamical states. The region occupied by the black dotted vertical lines highlight the range of standardised velocities, $0.3 < |\Delta\mathrm{V}|/\sigma_{r_{200}}\sim < 0.5$, which indicate an infaller population if their modal value lies within it. Each sub-population is normalised to their histogram densities, which is defined as $N = f_{i}/n(c_{i}-c_{i-1})$, where $f_{i}$ is the frequency per bin, $n$ is the total size of the histogram sample and $(c_{i}-c_{i-1})$ is the bin width.}
    \label{fig:bkspl}
\end{figure*}

AGN-hosting cluster galaxies are commonly found to coincide along the virialised boundaries of galaxy clusters and one explanation for this effect could potentially be that AGN sub-populations are backsplash galaxies, which are described as galaxies that have have already passed through their clustocentric pericentre on first infall and are now journeying towards their respective apocentres. 
Indeed, \cite{RomanOliveira2019} find that their more extreme jellyfish galaxies were more likely to lie along these boundaries, therefore, it is possible to consider that the AGN triggering could occur during the the pericentre passage and this activity continues as a (possibly) weaker AGN remnant of that journey until the activity is eventually quelled.
Therefore, in Figure \ref{fig:bkspl} we plot a series of $|\Delta\mathrm{V}|/\sigma_{r_{200}}$ histograms for our AGN and non-AGN sub-populations for each of the dynamical states at two radial bins for cluster galaxies $\leq r_{200}$ and those > $r_{200}$ (with the upper limit of 2.5 $r_{200}$), following the same procedures as \cite{Gill2005} and \cite{Pimbblet2011}.
These procedures involve noting the way in which infaller and backsplash galaxies could be defined. 
To elaborate, \cite{Gill2005} states that at $\sim R_{\mathrm{virial}}$ a population of cluster galaxies are infallers if they posses the mode value of $|\Delta\mathrm{V}|\approx400$ kms$^{-1}$.

For consistency, we adopt the translation of this to the absolute velocities of cluster galaxies normalised by their respective galaxy cluster velocity dispersions into the range $0.3 < |\Delta\mathrm{V}|/\sigma_{r_{200}} < 0.5$ as deduced by \cite{Pimbblet2011}.
Thus, if the mode of the standardised velocities for a sub-population has its foci at around $0.3 < |\Delta\mathrm{V}|/\sigma_{r_{200}} < 0.5$ for values around the virial radius, which we assume to be $R_{\mathrm{virial}}\sim r_{200}$, said sub-population would be classified as infalling.
In contrast, a sub-population of backsplash cluster galaxies would be expected to peak significantly at $|\Delta\mathrm{V}|/\sigma_{r_{200}}\sim0$ for values at or beyond our definition of the virial radius, with their fraction reaching zero at some upper limit (e.g. \citealt{Mamon2004,Pimbblet2011,Bahe2013,Haggar2020}).
Therefore, with respect to Figure \ref{fig:bkspl}, we see that the column of our non-merging sub-populations across both bins of radius do not show any significant difference in the distributions of velocities with the exception of those that lie $\leq r_{200}$, which show the non-AGN sub-population to occupy a mode within the range that nominally represents infallers, most likely for cluster galaxies $0.5 \leq r_{200} < 1.0$ \citep{Gill2005}.
Additionally, the AGN sub-population slightly deviates from the non-AGN velocity distribution with a mode centred at $|\Delta\mathrm{V}|/\sigma_{r_{200}}\sim0.8$, which could indicate stronger infalling.
In contrast the column of our merging AGN sub-populations show the strongest deviations from the distribution of non-AGN, especially with the $> r_{200}$ bin showing a significant centrally dominated AGN sub-population, where such a central dominance in relative velocity corresponds to a sub-population that were predominantly backsplash cluster galaxies.
However, the dependence of this being the true nature of the sub-population relies upon more precise definitions of the radii since there is a natural upper limit a bound cluster galaxy can extend outward to with respect to its galaxy cluster's potential, known as the splashback radius \citep{More2015,More2016}.
In addition, \cite{Haggar2020} shows that the fraction of backsplash galaxies diminishes by $2 r_{200}$ and $2.5 r_{200}$ for massive ($\sim\times10^{15}$M$_{\odot}$) merging and non-merging cluster systems respectively, thus demonstrating that merging cluster environments experience a greater decrease in the fraction of harbouring backsplash galaxies as one continues to extend beyond $r_{200}$.
Indeed, the sub-populations of the merging cluster galaxies present in the $\leq r_{200}$ bin show more variations in their general distributions with the modes of both the AGN and non-AGN sub-populations lying around $0.3 < |\Delta\mathrm{V}|/\sigma_{r_{200}} < 0.5$, which eludes to mostly infalling sub-populations rather than those associated with backsplash.
Finally, if one considers the equivalent peak of the AGN density histogram at $\Delta\mathrm{V}|/\sigma_{r_{200}}\sim1.7$ it could be possible there is a mix of recently accreted cluster galaxies and those that are relaxing onto a common potential. 
Although, it should be noted that not much information can be confidently derived from the AGN sub-populations within the bins that possess small samples size ($N\lesssim100$), especially with the merging AGN-hosting cluster galaxies at $\leq r_{200}$ that only has $N=15$.

\subsection{Phase-Space Analysis}
\label{ssec:rhee}

\begin{figure*}
    \centering
    \includegraphics[width=0.8\paperwidth]{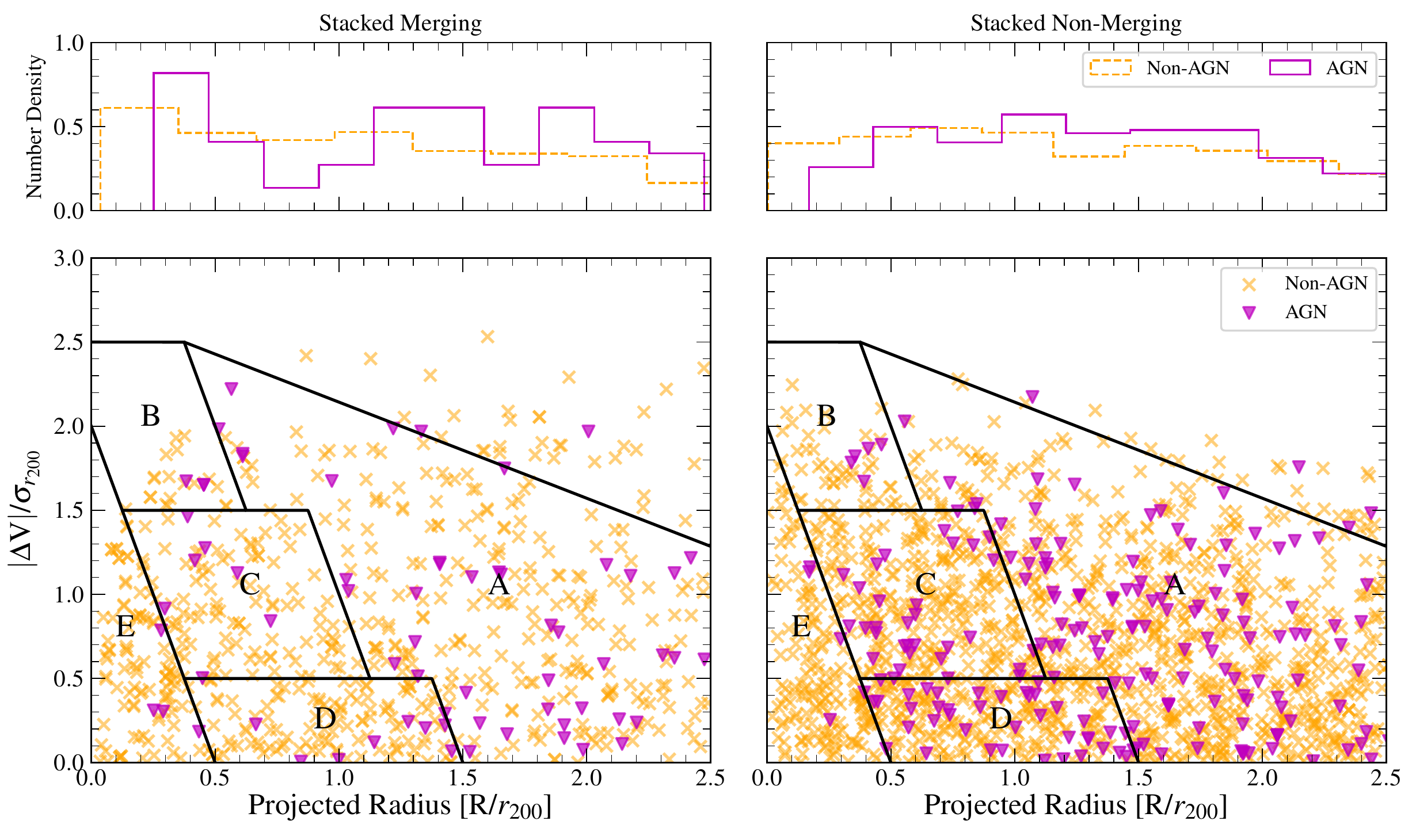}
    \caption{The merging (left column) and non-merging (right column) phase-space composites are depicted on the bottom row, where the absolute radial velocities are normalised as $|\Delta V|/\sigma_{r_{200}}$ and the projected radius as $R/r_{200}$. The regions A-E represent first, recent, recent-intermediate, intermediate and ancient infallers respectively as prescribed within Rhee et al. (2017). The corresponding histograms on the top row present the radii for each sub-population normalised to their respective histogram densities, which is defined as $N = f_{i}/n(c_{i}-c_{i-1})$, where $f_{i}$ is the frequency per bin, $n$ is the total size of the histogram sample and $(c_{i}-c_{i-1})$ is the bin width.}
    \label{fig:rhee}
\end{figure*}

In light of studying the modal absolute velocities between the core regions and the outer most radii for our composites in section \ref{ssec:bkspl} we attempt to make further sense of these distributions and their foci through a projected phase-space analysis.
To that end, we use the phase-space region analysis based on the $N$-body cosmological simulations of \cite{Rhee2017}.
Exploring the projected phase-space distributions of our cluster galaxy sub-populations for both dynamical states will allow for us to ascertain a cluster galaxy's time since first collapse onto the cluster potential and the likely stage of its journey at our current epoch of $z=0$.
Therefore, in Figure \ref{fig:rhee} we present the \cite{Rhee2017} projected phase-space for each dynamical state alongside the regions A-E, with each representing the space that is occupied by a cluster galaxy chronologically as it journeys through the cluster (e.g. first infall-coalesced onto potential).
As a complement to Figure \ref{fig:rhee}, we tabulate the numbers and fractions of the AGN sub-population for the merging and non-merging systems relative to each phase-space region in Table \ref{tab:rhee}.
The fractional uncertainties are computed from the $1\sigma$ confidence interval of a binomial distribution as analysed and depicted by \cite{Cameron2011}.

\begingroup
\renewcommand{\arraystretch}{1.5}
\begin{table}
    \centering
    \caption{The number and fraction of AGN for the merging and non-merging stacks within each phase-space region, as shown in Figure \ref{fig:rhee} with the lines of delineation originally defined by Rhee et al. (2017). The asymmetric uncertainties for each fraction represent the $1\sigma$ confidence interval of the binomial distribution (see Cameron 2011).}
    \begin{tabular}{ccccc}
        \hline
        Region & $N_{\mathrm{Merge}}$ & $f_{\mathrm{Merge}}$ & $N_{\mathrm{Non-Merge}}$ & $f_{\mathrm{Non-Merge}}$ \\
        \hline
        A & 42 & $0.15_{-0.02}^{+0.02}$ & 119 & $0.15_{-0.01}^{+0.01}$ \\
        
        B & 3 & $0.14_{-0.06}^{+0.09}$ & 5 & $0.09_{-0.04}^{+0.05}$ \\
        
        C & 7 & $0.07_{-0.02}^{+0.03}$ & 30 & $0.10_{-0.02}^{+0.02}$ \\
        
        D & 8 & $0.07_{-0.02}^{+0.03}$ & 43 & $0.16_{-0.02}^{+0.02}$ \\
        
        E & 4 & $0.04_{-0.02}^{+0.03}$ & 5 & $0.03_{-0.01}^{+0.02}$ \\
        \hline
    \end{tabular}
    \label{tab:rhee}
\end{table}
\endgroup

Purely by observation of Figure \ref{fig:rhee}, there is no obvious concentration of AGN in either dynamical state except by the overt imbalance between the sizes of each cluster sub-sample.
This is especially true for the co-added cluster galaxies that lie within the non-merging stack, which show a homogenised distribution of both sub-populations, although with the exception of an elevation of the AGN sub-population $1 \lesssim r_{200} \lesssim 2$.
However, the distribution of the radii in the merging stack highlights a peak at $0.25 \lesssim r_{200} \lesssim 0.50$ and cuts through segments of the post-accretion regions B-E.
Interestingly, the non-merging regions appear to show a `cut-off' along the line of delineation for the ancient infaller E region, with the exception of an insignificant number that do invade the region.

More importantly, one should contrast Figure \ref{fig:rhee} with the information in Table \ref{tab:rhee} to better interpret AGN concentration.
Thus, we note that region A has the most significant AGN contribution associated with first infallers, where both merging and non-merging stacks have a consistency between each other with $\sim15$ per cent across both sub-samples.
The merging composite maintains this fraction of AGN consistently into region B, albeit, tenuously so due to the greater uncertainties that do not significantly break away from the non-mergers combined with the difference in the number of galaxy clusters for each sub-sample.
Regions C and D are both considerably enriched for the non-merging composite comparatively against the merging composite with fractions of 10 and 16 per cent respectively.
Furthermore, in section \ref{ssec:whan} we determine the fractions of AGN in the merging and non-merging sub-samples to be 10.20 and 13.14 per cent respectively, which demonstrates an overall decrease in the total merging sub-sample AGN fraction.
Again, with reference to the different regions in Table \ref{tab:rhee}, it can be seen that the predominant source of this deficiency in merging cluster AGN fraction is in region D when taking into account the uncertainties and suggests AGN are somewhat quenched in merging cluster systems.
However, the discrepancy between the cluster sub-samples sizes does mitigate against this as a conclusive explanation for the differences in AGN fraction, especially when comparing clusters from each sub-sample individually in Table \ref{tab:bax}.
Additionally, it is estimated by \cite{Rhee2017} that the aforementioned backsplash galaxies would more commonly inhabit the regions C and D.
Therefore, implying these non-merging cluster AGN could have survived the first turnaround of their pericentres and potential quenching for up to $\lesssim3$Gyr post-turnaround depending on their distance to their apocentres.
Finally, the fractions of AGN-hosting cluster galaxies greatly diminish across both dynamical states in Region E, and this can be clearly seen in Figure \ref{fig:rhee} when contrasted with the non-AGN sub-populations suggesting AGN cannot survive, or are not commonly triggered, significantly in the ancient virialised regions of clusters.

\section{Discussion \& Summary}
\label{sec:conc}

The work we present here has the unfortunate discrepancy between our cluster sample sizes as a result of our implementation of the $\Delta$-test to enforce a significance to the 1 per cent level.
However, ensuring this strict criterion ensures we are selecting our substructured sub-sample to be a truer proxy of core merging processes and in spite of this we still have sufficient richness in the composites to make a comparative analysis.
Of course, the $\Delta$-test itself has its own misgivings operating as a proxy for core merging due to its reliance upon local deviations of cluster galaxies in $z$-space from the overall mean cluster values, alongside the projection effects due to the limitations of our 2D sky observations where we ultimately are unable to adequately resolve angular and radial separations .
Consequentially this results in a proxy of relatively recent cluster-cluster mergers that are in a late relaxing phase compared to systems with initial ICM interactions between two independent sub-clusters (e.g. see \citealt{Bulbul2016,Caglar2017}).
This leads us to ask the question, what do we mean by `merging'?
Merging clusters present processes with a variety of timescales dependent on the epoch of the merger and whether you observe the cluster galaxies or the ICM.
This is important when considering the origins of AGN themselves since they have been observed to be prevalent within `merging' systems as determined via ICM shock fronts \citep{Miller2003,Sobral2015}, as well as `Jellyfish galaxies' resulting from ram pressure stripping \citep{Owers2012,Ruggiero2019,Ebeling2019}, which could in turn be possible conspicuous tracers of AGN due to both being occasionally coincident (see \citealt{Poggianti2017,Marshall2018,RomanOliveira2019}).
Contrary to this however, it is shown that any minor merging processes indicated by the ICM do not have an immediate impact on the evolution of cluster galaxies \citep{Kleiner2014}.

Considering many clusters in our `non-merging' sample are actually exhibiting merging processes (e.g. see \citealt{Nulsen2013,Wen2013}) in the radio or X-ray implies we may not be capturing the true kinematic effects from AGN triggering due to ram pressure stripping activity, thus, an alternative way of determining merging galaxy clusters may be better suited.
In fact our AGN cluster galaxies in this work are optically selected, which therefore means our AGN sample contains the most efficient accretors.
To maintain such a high efficiency requires a consistent stream of cold gas funnelled from a sufficient reservoir, however, denser environments such as of that found towards the inner core regions of galaxy clusters ($\lesssim r_{200}$) do not typically yield such a supply.
In contrast, inefficiently accreting AGN may result from `drip-feeding' of the cold gas due to a variety of either in-situ or ex-situ processes (e.g. see \citealt{Hardcastle2007,Ellison2015}).
With this in mind the inefficient accretion onto the supermassive black hole could therefore be enough to power an AGN to provide signatures in the radio band, implying that radio selected AGN may provide a greater insight into the interplay between different modes of accretion; radio AGN with a low power output are commonly found in cluster galaxies that pervade the centres of galaxy clusters and groups \citep{Best2007,Ching2017}.
Our selection biasing of accretion efficient AGN can be seen in Figure \ref{fig:rhee} as the numbers depreciate as $r_{200} \rightarrow 0$, especially for cluster galaxies within region E, the slight increase in number for merging states is most likely the result of heavy interactions that displace or `throw' the cluster galaxies into different regions.
Contemplating on this further, we also applied a rather strict criteria to selecting our AGN using the WHAN diagram to maintain high significance in our emission lines while alleviating the loss in data that BPT diagrams would induce.
However, restricting our AGN selection to cluster galaxies having a strong EW$_{\mathrm{H}\alpha}$ $\geq$ 6\AA \ emission inevitably removes a sub-sample of weaker AGN that could possibly resemble a relatively ancient trigger in activity due to the local environment.
Although, the quid pro quo nature of relaxing this strict criteria would lead to contamination of emissions from AGB stars or LIER hosting cluster galaxies.

There is the additional possibility that our application of surface caustics to the cluster sample is too restrictive for those possessing merging environments leading to the omission of genuine members that are temporarily thrown out of the system before collapsing back onto the cluster.
However, there the cautious approach is often required to prevent lingerers from pervading the galaxy cluster membership for our sub-samples at the expense of potentially losing members in our merger.
Indeed, this is a problem that becomes more apparent for galaxy clusters in our sample in relative close proximity to other large, and independent, structures such as Abell 2065 which is currently undergoing merging processes with another cluster core \citep{Markevitch1999,Belsole2005,Chatzikos2006}; Abell 1750 is a part of a triple cluster system with ICM interactions that is $<1000$ kms$^{-1}$ from the central sub-cluster, risking overlapping cluster galaxies from these other structures due to our line-of-sight limitations \citep{Molnar2013,Bulbul2016}.

Within this work we have obtained a sample of 33 galaxy clusters collated with the BAX cluster database that were split into two sub-samples of 8 merging (relaxing) and 25 non-merging (relaxed) dynamical states from the $\Delta$-test for substructure \cite{Dressler1988}.
Compiling each of their memberships with MPA-JHU DR8 galaxies via the mass estimations methods of surface caustics \citep{Diaferio1997,Diaferio1999} sub-populations between AGN and non-AGN-hosting cluster galaxies were determined adhering to the strict criterion of log$_{10}$([\ion{N}{II}]/\ion{H}{$\alpha$})$\geq-0.32$ and EW$_{\mathrm{H}\alpha}$ $\geq$ 6\AA \ to the WHAN diagram \citep{CidFernandes2010,CidFernandes2011,Gordon2018a}.
This results in a kinematic analysis through the VDPs, rotational profiles and their respective positions in phase-space for each dynamical state.
The summary of our findings are as follows:

\begin{enumerate}[(i)]
   \item Merging cluster dynamical states on average, as determined by the $\Delta$-test, present kinematically active AGN within core regions ($< r_{200}$) that implies they are a first infaller and recently accreted sub-population of merging systems. This is coincident within regions where ram pressure is strongest for first pericentre passage (see \citealt{Ricarte2020}).
    
    \item Non-merging cluster dynamical states on average illustrate an AGN sub-population that is kinematically inactive and is homogenous with the non-AGN sub-population, with their VDPs being atypical for a relaxed galaxy cluster system, suggesting there is no unique behaviour that could infer mechanisms that affect AGN activity. 
    
    \item Phase-space analysis exhibits a fractional enrichment of AGN in non-merging cluster dynamical states in regions associated with `backsplash' cluster galaxies, which resemble galaxies that have made their first passage through their pericentre.
\end{enumerate}

\section*{Acknowledgements}

\noindent
We would like to extend our thanks to the anonymous referee for their comments and suggestions during peer review of this work.

\noindent
KAP acknowledges the support of STFC through the University of Hull's Consolidated Grant ST/R000840/1.
YAG is supported by the National Sciences and Engineering Research Council of Canada (NSERC).

\noindent
I would additionally like to thank Claire Lin for their unwavering support during the current SARS-CoV-2 (COVID-19) pandemic.

\noindent
This research made use of Astropy, a community-developed core Python package for Astronomy (\citealt{Collaboration2013}; \citealt{Collaboration2018}).

\noindent
This research has made use of the X-rays Clusters Database (BAX)
which is operated by the Laboratoire d'Astrophysique de Tarbes-Toulouse (LATT),
under contract with the Centre National d'Etudes Spatiales (CNES). 

\noindent
This research has made use of the NASA/IPAC Extragalactic Database (NED), which is operated by the Jet Propulsion Laboratory, California Institute of Technology, under contract with the National Aeronautics and Space Administration.

\noindent
Funding for SDSS-III has been provided by the Alfred P. Sloan Foundation, the Participating Institutions, the National Science Foundation, and the U.S. Department of Energy Office of Science. The SDSS-III web site is \url{http://www.sdss3.org/}.
SDSS-III is managed by the Astrophysical Research Consortium for the Participating Institutions of the SDSS-III Collaboration including the University of Arizona, the Brazilian Participation Group, Brookhaven National Laboratory, Carnegie Mellon University, University of Florida, the French Participation Group, the German Participation Group, Harvard University, the Instituto de Astrofisica de Canarias, the Michigan State/Notre Dame/JINA Participation Group, Johns Hopkins University, Lawrence Berkeley National Laboratory, Max Planck Institute for Astrophysics, Max Planck Institute for Extraterrestrial Physics, New Mexico State University, New York University, Ohio State University, Pennsylvania State University, University of Portsmouth, Princeton University, the Spanish Participation Group, University of Tokyo, University of Utah, Vanderbilt University, University of Virginia, University of Washington, and Yale University.


\section*{Data Availability}

The SDSS DR8 and MPA-JHU data utilised within this work are publicly available to examine and download from the SQL-based SkyServer CasJobs service, which can be accessed via \url{http://skyserver.sdss.org/CasJobs}.




\bibliographystyle{mnras}
\bibliography{lebthesis.bib} 








\bsp	
\label{lastpage}
\end{document}